\documentclass[prl,twocolumn,aps,superscriptaddress]{revtex4-1}
\usepackage{graphicx,graphics,color,epsfig}
\usepackage{times}
\usepackage{epstopdf}
\usepackage{bm}
\usepackage{amsmath}
\usepackage{amssymb}
\usepackage{color}
\usepackage{diagbox}
\usepackage{appendix}
\usepackage{dcolumn}
\usepackage[colorlinks=true, letterpaper=true, pdfstartview=FitV, linkcolor=blue, citecolor=blue, urlcolor=blue]{hyperref}

\begin{document}

\title{Chiral Phonons in Chiral Materials}

\author{Hao Chen}
\affiliation{NNU-SULI Thermal Energy Research Center (NSTER) \& Center for Quantum Transport and Thermal Energy Science (CQTES), School of Physics and Technology, Nanjing Normal University, Nanjing 210023, China}
\affiliation{Division of Physics and Applied Physics, School of Physical and Mathematical Sciences, Nanyang Technological University, Singapore 637371, Singapore}

\author{Weikang Wu}
\affiliation{Division of Physics and Applied Physics, School of Physical and Mathematical Sciences, Nanyang Technological University, Singapore 637371, Singapore}
\affiliation{Research Laboratory for Quantum Materials, Singapore University of Technology and Design, Singapore 487372, Singapore }

\author{Jiaojiao Zhu}
\affiliation{Research Laboratory for Quantum Materials, Singapore University of Technology and Design, Singapore 487372, Singapore }

\author{Weikang Gong}
\affiliation{College of Life Science and Chemistry, Faculty of Environmental and Life Sciences, Beijing University of Technology, Beijing 100124, China}
\affiliation{Division of Mathematical Sciences, School of Physical and Mathematical Sciences, Nanyang Technological University, Singapore 637371, Singapore}

\author{Weibo Gao}
\email{wbgao@ntu.edu.sg}
\affiliation{Division of Physics and Applied Physics, School of Physical and Mathematical Sciences, Nanyang Technological University, Singapore 637371, Singapore}

\author{Shengyuan A. Yang}
\email{shengyuan\_yang@sutd.edu.sg}
\affiliation{Research Laboratory for Quantum Materials, Singapore University of Technology and Design, Singapore 487372, Singapore }

\author{Lifa Zhang}
\email{phyzlf@njnu.edu.cn}
\affiliation{NNU-SULI Thermal Energy Research Center (NSTER) \& Center for Quantum Transport and Thermal Energy Science (CQTES), School of Physics and Technology, Nanjing Normal University, Nanjing 210023, China}

\begin{abstract}

The concept of chirality makes ubiquitous appearance in nature. Particularly, both a structure and its collective excitations may acquire well defined chiralities. In this work, we reveal an intrinsic connection between the chiralities of a crystal structure and its phonon excitations. 
We show that the phonon chirality and its propagation direction are strongly coupled with the lattice chirality, which will be reversed when a chiral lattice is switched to its enantiomorph. In addition, distinct from achiral lattices, propagating chiral phonons exist for chiral crystals also on the principal axis through the $\Gamma$ point, which strengthens its relevance to various physical processes. We demonstrate our theory with a 1D helix-chain model and with a concrete and important 3D material, the $\alpha$-quartz. We predict a chirality diode effect in these systems, namely, at certain frequency window, a chiral signal can only pass the system in one way but not the other, specified by the system chirality. Experimental setups to test our theory are proposed. Our work discovers fundamental physics of chirality coupling between different levels of a system, and the predicted effects will provide a new way to control thermal transport and design information devices.

\end{abstract}

\maketitle

\textcolor{blue}{\textit{Introduction} ---}
The concept of chirality plays a significant role in multiple branches of sciences such as physics, chemistry~\cite{prelog1976chirality,cahn1966specification,francotte2006chirality}, and biology. An object or a structure is chiral if it is distinguished from its mirror image. In other words, all possible mirror (inversion) symmetries of the structure must be broken. Such chiral structures are ubiquitous, ranging from microscopic molecular structures to astronomical objects, and they have found wide applications, such as in enantioselective catalysis~\cite{ma2009enantioselective}, chiral drug design~\cite{reddy2004chirality}, chiral-induced spin selectivity~\cite{gohler2011spin,guo2012spin,liu2021chirality}, and etc.

Chirality also applies to the description of collective excitations of a system. A notable example from recent research is the chiral phonon~\cite{zhang2015chiral}---the collective oscillation modes of a crystal lattice with a definite handedness. It has been  shown that chiral phonons can couple with circularly polarized light and valley electrons and result in chiral selective optical transitions~\cite{zhu2018observation,chen2019entanglement,li2019momentum,he2020valley}. And it was proposed that chiral phonons can make important contributions to the thermal conductivity~\cite{pandey2018symmetry}, phononic Hall effect~\cite{grissonnanche2020chiral,park2020phonon}, orbital magnetization~\cite{hamada2018phonon,juraschek2019orbital,cheng2020large}, and anomalous thermal expansion~\cite{romao2019anomalous}. Most recently, the concept of chiral phonons has been extended to three-dimensional (3D) systems, enabling propagating chiral phonons~\cite{chen2021propagating}.

Now, a natural question is: \emph{Is there any interplay between chiral phonons and the chirality of the underlying lattice?} This important question has remained unexplored and studies so far mainly focused on achiral structures. Intuitively, for an achiral system, each chiral phonon mode must have a chiral partner connected by a mirror operation, which may pose difficulty to single out one chirality. For example, each left-handed propagating chiral phonon mode in Ref.~\cite{chen2021propagating} has a right-handed partner that propagates with the same velocity, so that the detection would require a complicated scheme involving valley scattering~\cite{zhu2018observation,chen2021propagating}.

In this work, we address the above question by exploring chiral phonons in chiral lattices. We reveal a strong coupling between the chiralities of the two. Distinct from achiral structures, we find that a propagating chiral phonon mode has its propagation direction tied to the lattice chirality. At a fixed frequency, phonons with opposite chiralities propagate in opposite directions. The phonon chirality or the propagation direction is switched when the chiral lattice is changed to its enantiomorph. Furthermore, for a chiral crystal, chiral phonons can appear on the principal axis that goes through the $\Gamma$ point, such that we can excite them directly without the intervalley process. These results are confirmed via explicit calculations on a helix-chain model and demonstrated in a common and important 3D material --- $\alpha$-quartz. We predict a chirality diode effect in these systems, i.e., the input chirality information (e.g., by a circularly polarized light) can only pass the system via chiral phonons in one way but not the other, as specified by the system chirality. Possible experimental detections of our proposed effects are discussed. Our work discovers a fundamental connection between the chiralities of a structure and its excitations, which opens a new direction for phononic research. The predicted chirality diode effect may enable new routes for controlling thermal transport and for designing information devices.

\begin{figure}[tbp]
	\includegraphics[width=2.8 in,angle=0]{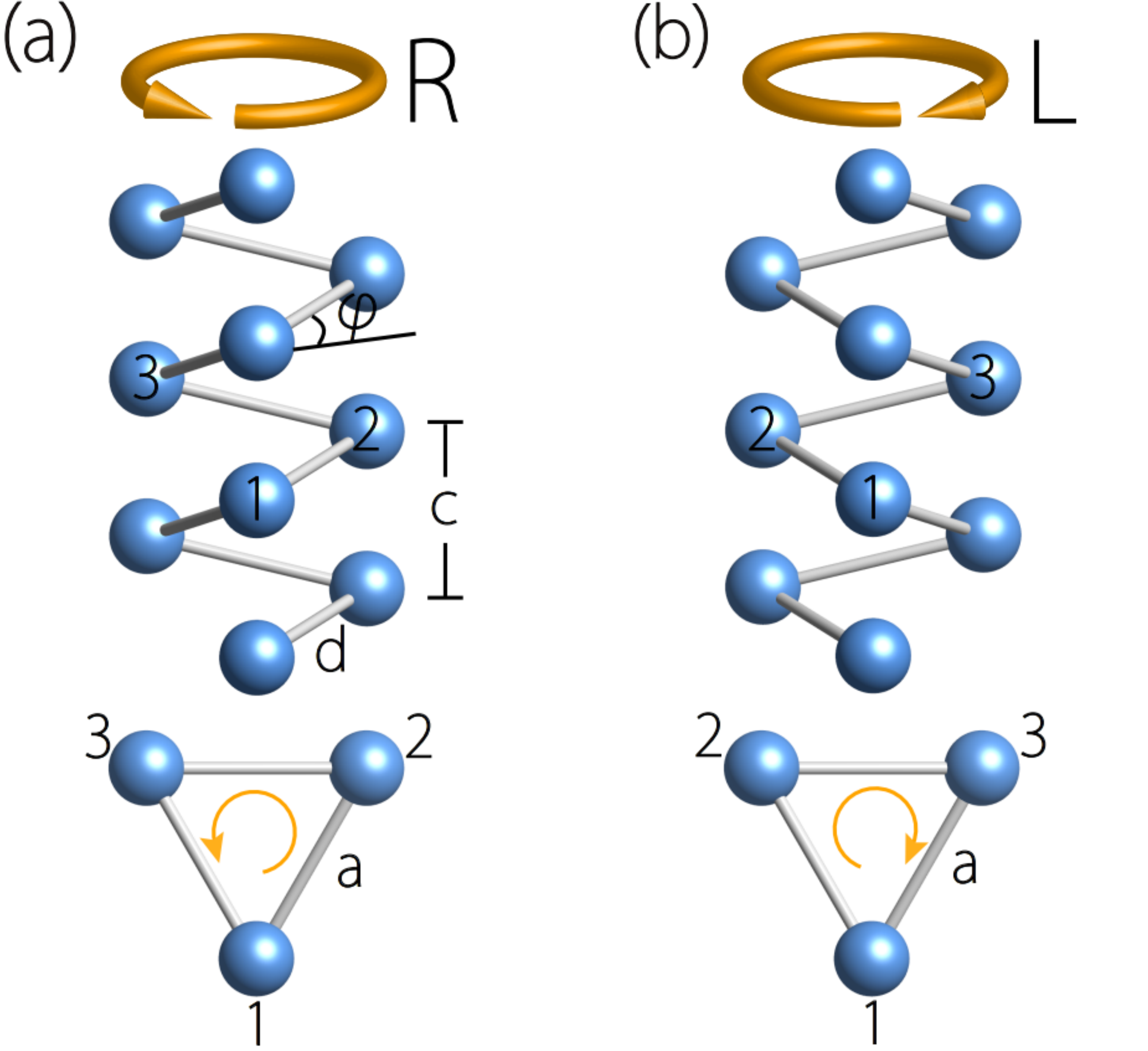}
	\caption{\label{fig1} (a) right-handed and (b) left-handed 1D helix chain model. The bottom panel shows the top view. The structural parameters are marked in (a).}
\end{figure}

\textcolor{blue}{\textit{Helix-chain model} ---}
To capture the essential physics, we first consider the simplest chiral lattice model, a 1D helix chain. As illustrated in Fig.~\ref{fig1}, the simplest helix has three sites in a unit cell and its handedness can be readily determined from its spiral pattern: Fig.~\ref{fig1}(a) and Fig.~\ref{fig1}(b) show right and left handed chains, respectively. Clearly, the helix structure breaks all possible mirrors, yet for a regular helix, it preserves a threefold screw axis $S_{3z}=\{C_{3z}|00\frac{1}{3}\}$ along the chain ($z$ direction).
In the ground state (i.e., without oscillation), the chiral structure can be specified by a few parameters, as indicated in Fig.~\ref{fig1}(a). Here, the top view is a equilateral triangle with side length $a$, $d$ is the bond length between two neighboring sites, $\varphi=\arccos (a/d)$ is the tilting angle of a bond from the $x$-$y$ plane, and the helix period $c=3a \tan \varphi$.

The harmonic oscillations of the lattice is described by the standard Hamiltonian
\begin{equation}\label{Hph}
	\mathcal{H}=\frac{1}{2}p^T p + \frac{1}{2}u^T K u,
\end{equation}
where $u$ is a column vector of displacements from the equilibrium positions, multiplied by the square root of mass (the mass is taken to be the same for every site here); $p$ is the conjugate momentum vector, and $K$ is the force constant matrix. For simplicity, we retain only the nearest-neighbor coupling. These parameters, along with the structural parameters such as $\varphi$, specify the matrix $K$. The phonon spectrum $\omega_{k,\sigma}$ and the phonon eigenmodes $\epsilon_{k,\sigma}$ are then solved from the equation
\begin{equation}\label{eqs_dymtrp}
	D(k)\epsilon_{k,\sigma}=\omega_{k,\sigma}^{2}\epsilon_{k,\sigma},
\end{equation}
where $k$ is the 1D wave vector along the chain, the dynamic matrix $D(k)$ is the spatial Fourier transform of $K$, and the index $\sigma$ labels the phonon branch.

\begin{figure}[tbp]
	\includegraphics[width=1\linewidth]{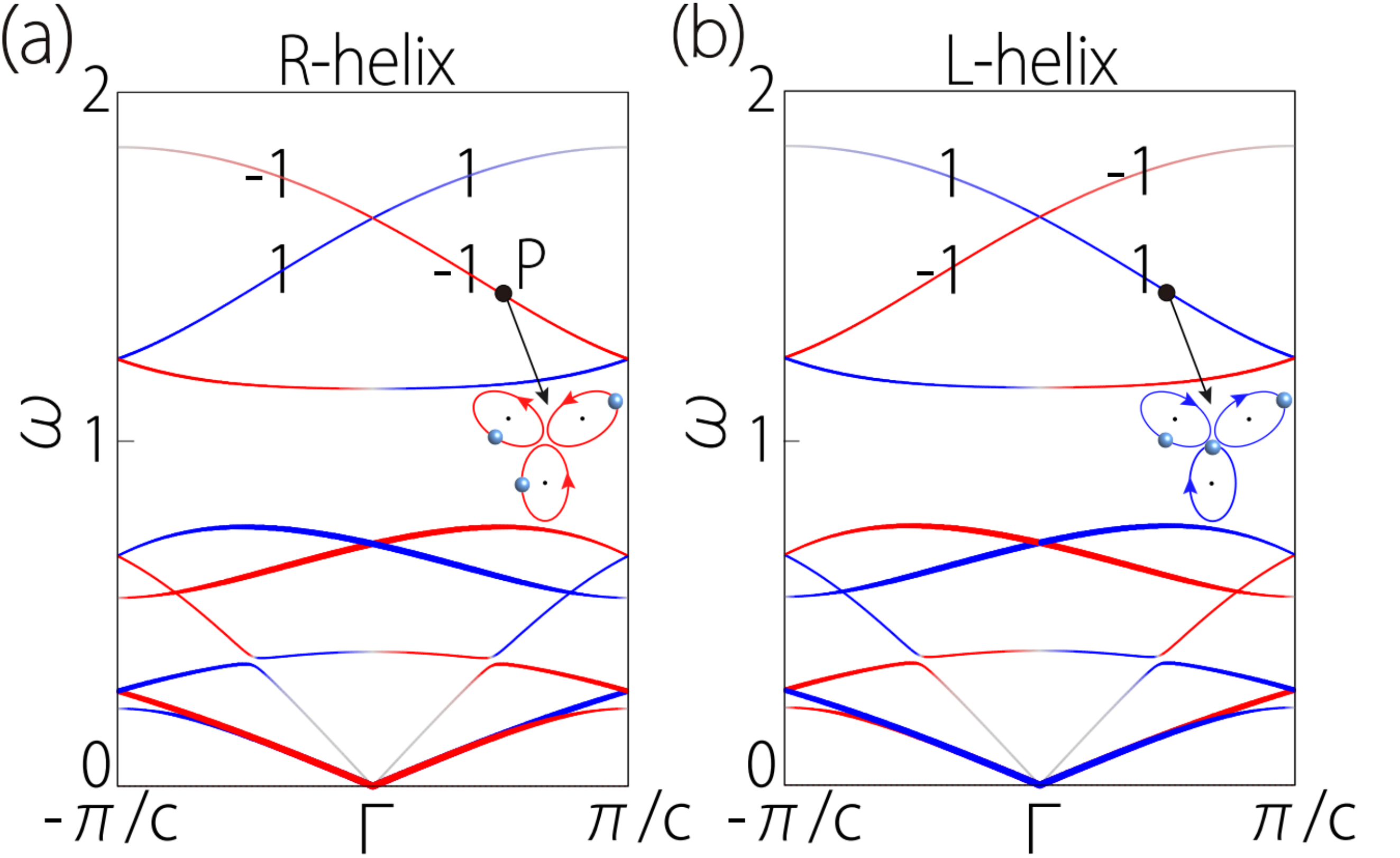}
	\caption{\label{fig2} 
Calculated phonon spectra for (a) right-handed helix and (b) left-handed helix. They show the same dispersion but opposite chirality distribution. The red and blue colors represent the right and left circularly polarized phonon modes, respectively. The PAM for the top two phonon branches are marked. The inset shows the top view of the phonon vibration pattern for the mode marked by the black dot. The calculation details and model parameters are presented in the Supplemental Material~\cite{S.M}.}
\end{figure}

The calculated phonon spectra of right- and left-handed helix are plotted in Fig.~\ref{fig2}(a) and (b), which consist of three acoustic branches and six optical branches, corresponding to the three sites in a unit cell. Particularly, to investigate the chirality of phonon modes, we use the color map in Fig.~\ref{fig2} to indicate the value of phonon circular polarization along the chain, which is given by~\cite{zhang2015chiral}
\begin{eqnarray}\label{eq:polarization}
	s_{k,\sigma} = \epsilon^{\dag}_{k,\sigma}\hat{S}_z\epsilon_{k,\sigma}
\end{eqnarray}
where
\begin{equation}
  \hat{S}_z=\sum_{\alpha=1}^{3}(|R_{\alpha}\rangle\langle R_{\alpha}| - |L_{\alpha}\rangle\langle L_{\alpha}|)
\end{equation}
is the circular polarization operator, $|R_\alpha\rangle$ ($|L_\alpha\rangle$) is the right (left) circularly polarized oscillation basis at site $\alpha$ (for circular motion in the $x$-$y$ plane). It was shown in Ref.~\cite{zhang2015chiral} that $\hbar s_{k,\sigma} $ gives the angular momentum of the phonon mode
$\epsilon_{k,\sigma}$ along the chain. And the sign of $s_{k,\sigma}$ indicates the chirality, namely, the phonon mode is right (left) handed if $s_{k,\sigma}>0$ $(<0)$.

From the results in Fig.~\ref{fig2}(a), first of all, one can clearly see that the phonon modes are chiral. The inset shows the vibration pattern of the mode $P$ indicated by the black dot in Fig.~\ref{fig2}(a). One observes that from top view, all three sites make counter-clockwise rotations, confirming its right-handed chirality. This is in contrast with an achiral chain. For example, if the chain preserves a mirror (denoted by $M$) that contains the chain axis, then each mode $\epsilon_{k,\sigma}$ would be degenerate with a mirror partner $M \epsilon_{k,\sigma}$, such that there is no net circular polarization nor chirality of the phonons.

\begin{figure}[tbp]
	\includegraphics[width=1\linewidth]{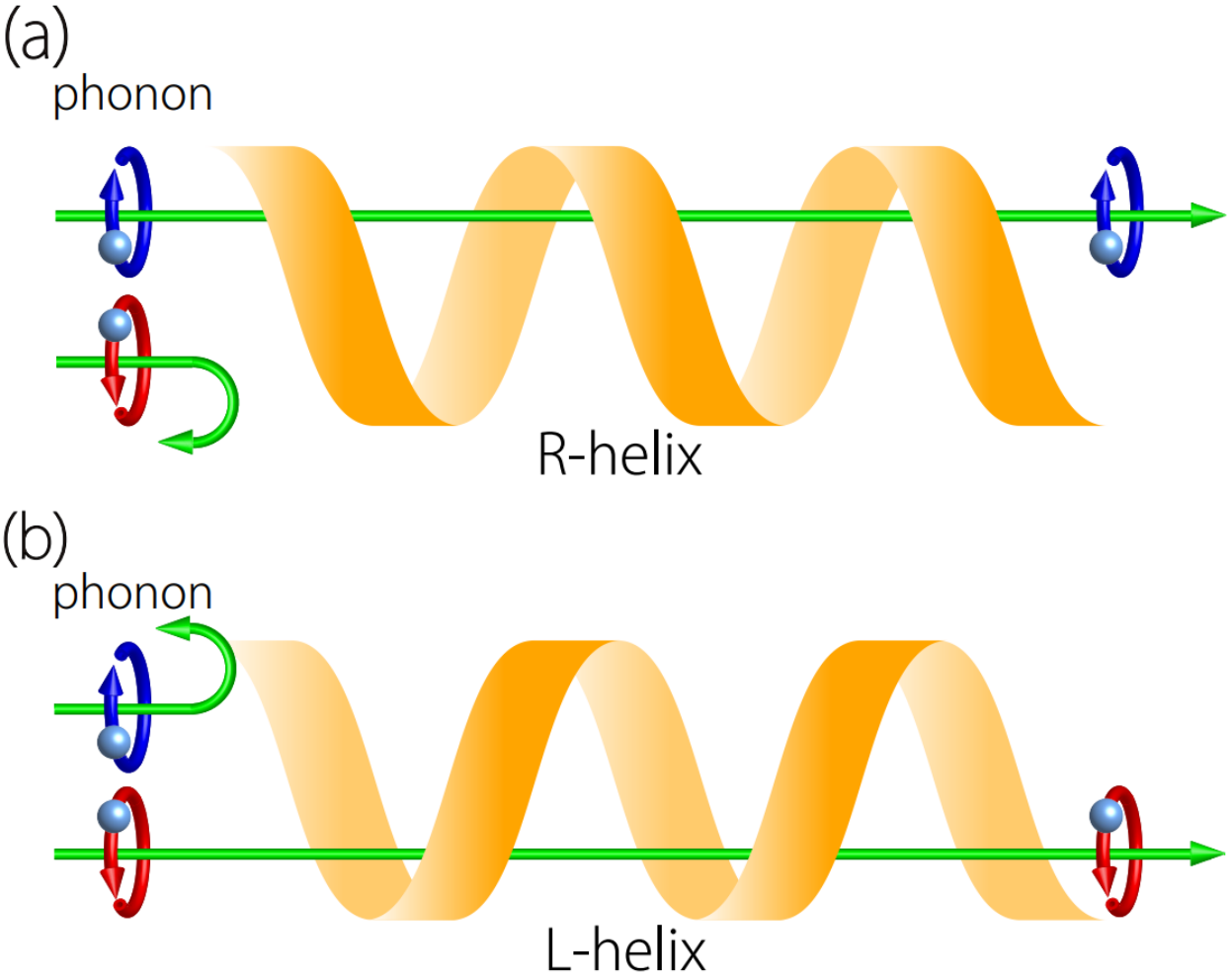}
	\caption{\label{fig3} Schematic diagram of the chirality filtering. (a) At the frequency around $P$ in Fig.~2(a), only left-handed phonons are allowed to pass the helix from left to right. (b) The situation reverses when the chirality of the helix is switched.}
\end{figure}

Second, the chiral phonons can generally propagate along the chain. The propagation direction is determined by the sign of the slope of the dispersion. This may be regarded as a kind of chirality-orbit coupling, i.e., the chirality of  phonon is coupled with its propagation. Importantly, there exists frequency windows in which the chirality is \emph{completely locked} with the propagation direction. For example, consider the frequency at mode $P$ in Fig.~\ref{fig2}(a). At this frequency, there are only two phonon modes. The one that propagates in the $+z$ direction is left-handed, whereas that propagates in the $-z$ direction is right-handed, as dictated by the time reversal symmetry. Thus, the helix chain acts as \emph{a chirality filter} for phonons when operated at this frequency. As schematically shown in Fig.~\ref{fig3}(a), the system allows the left-handed phonon to pass through from left to right, but not the right-handed one; and the situation is reversed for propagation in the opposite direction.

Third, as mentioned, the helix chain here preserves a threefold screw rotational symmetry $S_{3z}$. It allows the definition of a pseudo-angular momentum (PAM)~\cite{zhang2015chiral} $\ell_{k,\sigma}$ for phonons, which just corresponds to the eigenvalues of a phonon mode under threefold rotation (the fractional translation part can be disregarded as it does not affect any selection rules to be discussed later). Explicitly, we have
\begin{equation}\label{PAM}
	\mathcal{R}_z[(2\pi/3)]\epsilon_{k,\sigma}= e^{-i(2\pi/3)\ell_{k,\sigma}} \epsilon_{k,\sigma},
\end{equation}
where $\mathcal{R}_z$ is the rotation operator acting on the eigenmode and is defined to connect the three sites, and the PAM defined in this way is an integer $\in\{0,\pm 1\}$. For example, the PAM for the top two phonon branches are labeled in Fig.~\ref{fig2}(a). One observes that PAM is also coupled with the chirality and the propagation direction. Phonons with PAM $+1$ travel in the $+z$ direction, whereas those with $-1$ go the opposite way.

Finally, it is crucial and evident that the phonon chirality (and the PAM) is tied with the chirality of the lattice. The results above are for the right-handed helix. For a left-handed helix, the phonon chirality would simply be flipped, as in Fig.~\ref{fig2}(b). And the system allows the right-handed phonon to pass through from left to right, but not the left-handed one, as shown in Fig.~\ref{fig3}(b). This also justifies that for the achiral structure, which may be regarded as the phase boundary between the right and the left handed helices, the phonon chirality as well as the associated effects such as chirality-orbit coupling and chirality filtering must vanish.

\begin{figure*}[tbp]
	\includegraphics[width=1\linewidth]{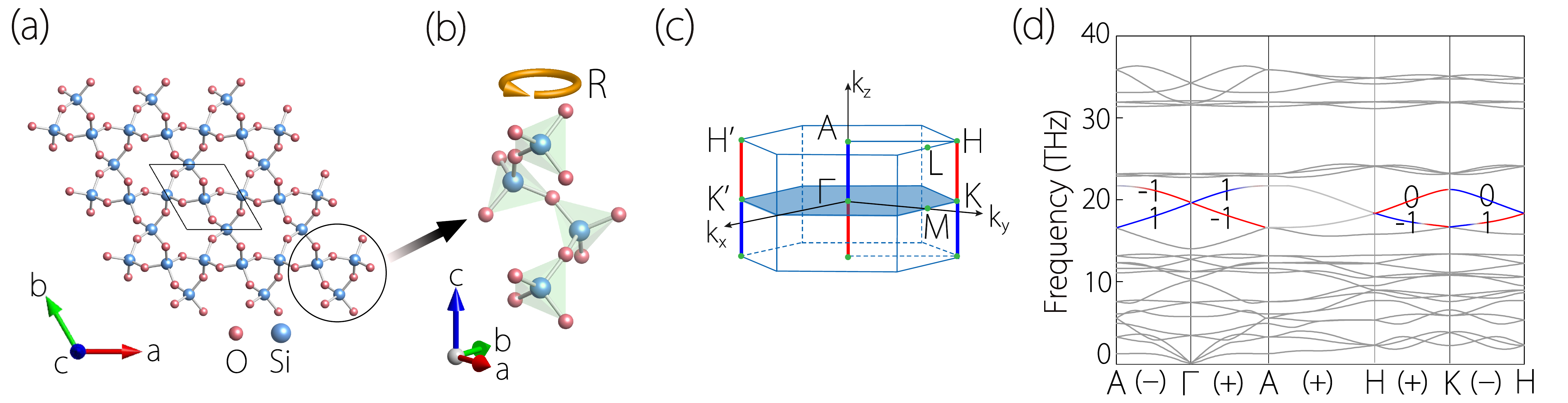}
	\caption{\label{fig4} (a) Top view of the right-handed $\alpha$-quartz (SiO$_2$), the black box marks the unit cell. (b) Side view of the vertical chain structure marked by the black circle in (a). It shows a right-handed helix structure. (c) Brillouin zone of the right-handed $\alpha$-quartz, the red and blue colors on the high-symmetry paths indicate the opposite distribution of phonon chirality in the frequency window around 20 THz. (d) Phonon spectrum of the right-handed $\alpha$-quartz. $+/-$ on the path indicate the $k_z>0$ $/$ $k_z<0$ parts. For the two phonon branches with energy in the range of 16$-$22 THz, we marked the the phonon chirality and PAM. Red/blue color represents right/left-handed phonons.}
\end{figure*}

\textcolor{blue}{\textit{$\alpha$-quartz: A concrete example} ---}
The physics we learned from the 1D helix model is completely general. Below, we extend the discussion to 3D crystal materials via a concrete example. Let's consider the material $\alpha$-quartz, which is one of the stable crystal polymorphs of SiO$_2$ that is most commonly found in nature~\cite{deal2013physics}. It has important applications in many industrial branches, ranging from construction to electronics. The lattice structure of quartz consists of SiO$_4$ tetrahedra linked by shared corner oxygens. Importantly, the $\alpha$-quartz lattice is chiral. It crystalizes in two enantiomorphic space groups, $P3_1 21$ (No. 152) and $P3_2 21$ (No. 154)~\cite{glinnemann1992crystal,antao2008state}, depending on the chirality. Figure.~\ref{fig4} (a) and (b) show the structure of the right-handed $\alpha$-quartz with space group $P3_1 21$. One observes that along the $c$ axis ($z$ direction), the SiO$_4$ tetrahedra spiral upwards in a right-handed way, just resembling the helix model in Fig.~\ref{fig1}(a). For the left-handed $\alpha$-quartz (not shown here), the spiral direction is reversed. In terms of space groups, both structures preserve a threefold screw axis, but the screw directions are opposite.

In Fig.~\ref{fig4}(d), we plot the phonon spectrum of right-handed $\alpha$-quartz obtained from first-principles calculations. The calculation details are given in the Supplemental Material~\cite{S.M}. There are totally twenty-seven phonon branches, corresponding to the nine atoms in the unit cell. From the calculated phonon mode eigenvectors, we can evaluate the phonon circular polarization $s_{\bm k, \sigma}$ just as what we did for the helix model. We confirm that the phonons here are generally chiral.

We are particularly interested in the phonons on the high symmetry paths that preserve the screw rotation along $c$, because these modes further allow a well defined PAM. For $\alpha$-quartz, these paths include $\Gamma$-$A$, $K$-$H$, and $K'$-$H'$ in the Brillouin zone (Fig.~\ref{fig4}(c)). Note that each path has two sections: the section with $k_z>0$ and the one with $k_z<0$. To distinguish the two, we shall label the former/latter one by adding a plus/minus sign. For instance, the two sections along the $k_z$ axis will be labeled as $\Gamma$-$A(+)$ and $\Gamma$-$A(-)$, respectively. In Fig.~\ref{fig4}(d), we see that the seventeenth and eighteenth phonon branches around 16--22 THz are well separated from other branches, which gives advantage for their excitation and detection. Focusing on these two branches, in Fig.~\ref{fig4}(d), we use the color map to show the calculated phonon circular polarization on the high-symmetry paths. Clearly, one observes that these phonons are chiral. For instance, the phonons of the eighteenth branch on the $\Gamma$-$A(+)$ are left-handed and propagate in the $+z$ direction, whereas the corresponding ones on $\Gamma$-$A(-)$ are right-handed and propagate in the $-z$ direction. In other words, the phonon chirality is coupled with the propagation direction.  This confirms the features of chirality-orbit coupling and chirality filtering which we learned from the helix model.

It is important to note that the emergence of chiral phonons on the $\Gamma$-$A$ path is a manifestation of the chirality of the system. For 3D achiral crystals, such as WN$_2$ discussed in Ref.~\cite{chen2021propagating}, phonon modes on this path do not have net chirality, since each left-handed mode must be accompanied with a right-handed partner. Meanwhile, the property of chirality filtering is generally not expected for an achiral crystal: a branch can have the same chirality on $K$-$H(+)$ and $K$-$H(-)$, as shown in Ref.~\cite{chen2021propagating}.

In Fig.~\ref{fig4}(d), we have also labeled the phonon PAM for the seventeenth and eighteenth branches. One can see that the PAMs for $\Gamma$-$A(+)$ and $\Gamma$-$A(-)$ also differ by a minus sign, as dictated by the time reversal symmetry. The PAM determines selection rules for phonon coupling with other particles. For example, a circularly polarized light incident along the $z$ direction can excite a phonon mode subjected to the PAM conservation: $m=\ell$ mod 3, where $m=\pm 1$ for right/left circularly polarized light. This allows a selective coupling between chiral phonons and infrared light.

The results above are for the right-handed $\alpha$-quartz. Evidently, for the left-handed crystal, the phonon spectrum would look the same but the phonon chirality and PAM would be flipped.

\textcolor{blue}{\textit{Discussion} ---}
We have discovered an intrinsic connection between the chiralities of a structure and its phonon excitations. We find that chiral structures host chiral phonons on the principal axis through $\Gamma$ which exhibit chirality-orbit coupling, chirality filtering, and net PAM, all distinct from achiral systems, and these properties are reversed when the structural chirality is switched.

\begin{figure}[tbp]
	\includegraphics[width=1\linewidth]{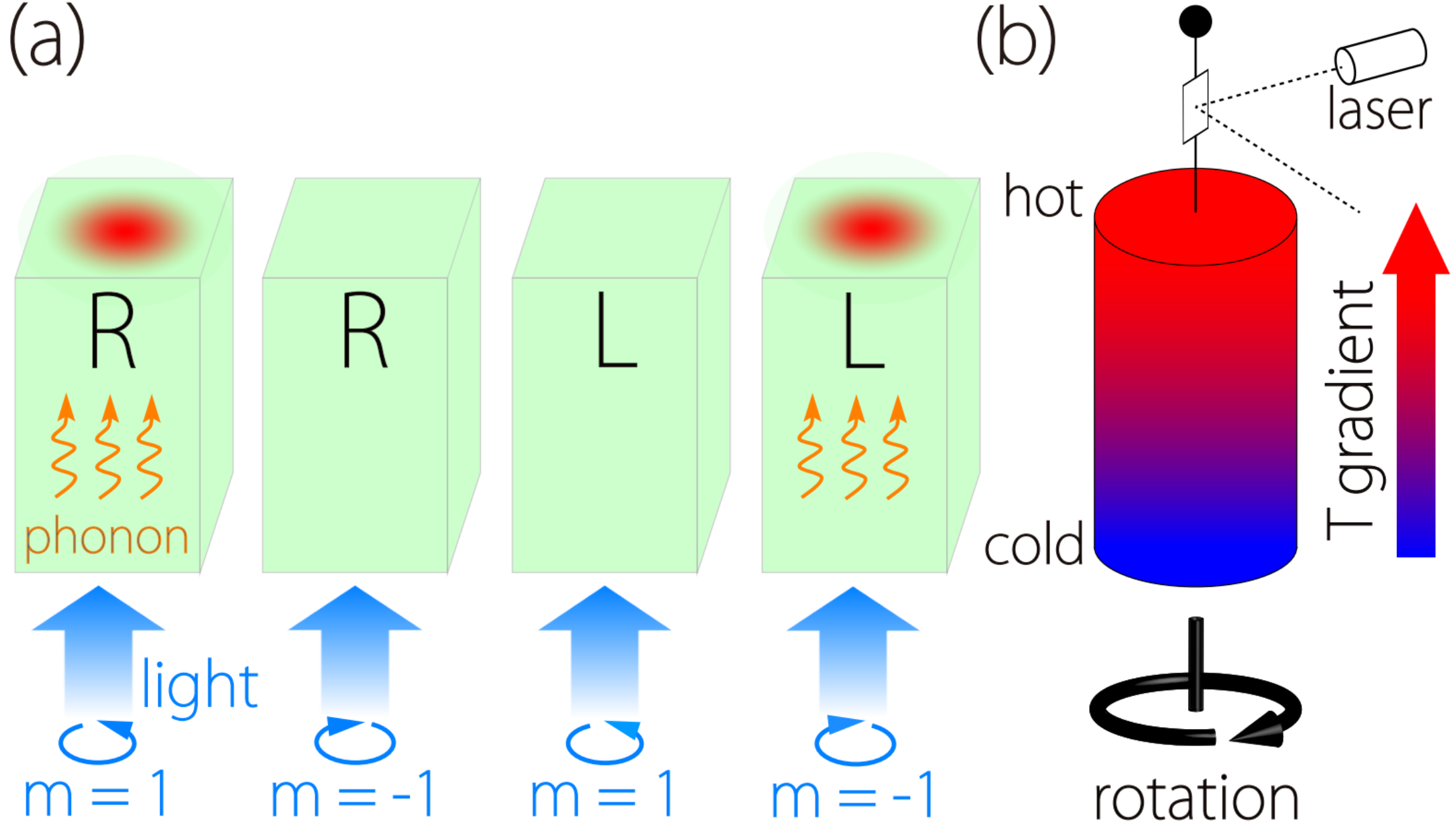}
	\caption{\label{fig5}  (a) Schematic experimental setup for detecting chirality diode effect. R and L on the sample indicate right-handed and left-handed lattices, respectively. The bottom of the sample is shined with a circular polarized light. The chirality diode effect manifests as the temperature difference on the top surface depending on the light polarization and the lattice chirality. (b) When a hanged $\alpha$-quartz sample is applied with a temperature gradient. The chiral phonons will induce a rotation of the sample and the rotation direction depends on the crystal chirality. See the main text for more descriptions.}
\end{figure}

The chirality filtering feature can lead to a chirality diode effect for phonons. For example, consider a slab of a chiral crystal, e.g., of right-handed $\alpha$-quartz, with its $c$ axis along the $z$ direction. We apply a circularly polarized infrared light with frequency peaked around 20 THz (i.e., the range of eighteenth branches in Fig.~\ref{fig4}(d)) incident on the bottom of the slab, as shown in Fig.~\ref{fig5}(a). Due to the PAM conservation, a right/left circularly polarized light can resonantly excite phonon modes with PAM $+1$/$-1$. Meanwhile, due to momentum conservation, light can only excite phonons on the $\Gamma$-$A$ paths. (Note that the slab surface breaks the translation symmetry along $z$, so $k_z$ needs not be conserved.) Therefore, the incident light with right (left) circular polarization will mostly excite chiral phonons with PAM $+1$($-1$) on the $\Gamma$-$A$ path. However, one observes that in this frequency range, only the phonons with PAM $+1$ are propagating in the $+z$ direction, i.e., can pass through the slab. This makes the system act like a diode for chiral phonon propagation. Since phonons are important for heat conduction (electronic contributions  here can be neglected as $\alpha$-quartz are good insulator with a band gap $\sim$ 5.8 eV), in experiment, by switching the circular polarization of the incident light, the chirality diode effect can manifest as a temperature difference detected on the top surface of the slab, as illustrated in Fig.~\ref{fig5}(a).

Due to the chirality-orbit coupling, when a temperature gradient along the $c$ axis is applied across the sample, the non-equilibrium phonon distribution $f_{\bm k,\sigma}$ will generally cause a nonzero net phonon circular polarization $S=\sum_{\bm k,\sigma}s_{\bm k,\sigma} f_{\bm k,\sigma}$ and hence a nonzero phonon angular momentum $\hbar S$~\cite{zhang2014angular,hamada2018phonon}. This thermally generated phonon angular momentum can be detected by the rotation of a hanged cylinder shaped sample similar to the Einstein-de Haas effect~\cite{frenkel1979history}, as shown in Fig.~\ref{fig5}(b). From our calculations, the resulting angular velocity for $\alpha$-quartz is $\omega\sim \frac{\Delta T/(1\text{K})}{hr^2/(1\text{m})^3}\times 10^{-20}\ \text{s}^{-1}$~\cite{S.M}. Assuming that the radius and the height of the sample are $r=10\ \mu \text{m}$ and $h=100\ \mu \text{m}$, and the temperature difference between the top and the bottom surfaces $\Delta T=10 \ \text{K}$, then the resulting $\omega$ can reach $\sim 10^{-5}\ \text{s}^{-1}$, which can be detected in experiment. In addition, for ionic crystals, the chiral phonons may also carry a magnetic moment. Then the net phonon angular momentum is accompanied with an induced magnetization, which can be detected by magnetism probes. Again, all these signals are tied with the chirality of the crystal: the signals should switch sign when we change a left-handed sample with a right-handed one.

\textcolor{blue}{\textit{Aknowledgment} ---}
The authors thank D. L. Deng for helpful discussions.
This work was supported by NSFC (11890703, 11975125) and Singapore MOE AcRF Tier 2 (MOE2019-T2-1-001). Hao Chen was also supported by China Scholarship Council No. 202006860020. We acknowledge computational support from the Texas Advanced Computing Center.

\bibliography{aps-main}

\end{document}